\documentclass[amsthm]{elsart}

\usepackage{yjsco}
\usepackage{natbib}
\usepackage{hhline}

\begin{document}
\begin{frontmatter}

\title{Performance of Buchberger's Improved Algorithm using Prime Based Ordering}

\thanks{This research was partly supported by the School of Engineering and Information Technology, Deakin University}

\author{Peter Horan and John Carminati}
\address{Deakin University, Geelong, Australia, 3217}
\ead{peter@deakin.edu.au, jcarm@deakin.edu.au}


\begin{abstract}
Prime-based ordering which is proved to be admissible, is the encoding of indeterminates in power-products with prime numbers and ordering them by using the natural number order. Using Eiffel, four versions of Buchberger's improved algorithm for obtaining Gr\"obner Bases have been developed: two total degree versions, representing power products as strings and the other two as integers based on prime-based ordering. The versions are further distinguished by implementing coefficients as 64-bit integers and as multiple-precision integers. By using prime-based power product coding, iterative or recursive operations on power products are replaced with integer operations. 
It is found that on a series of example polynomial sets, significant reductions in computation time of 30\% or more are almost always obtained.
\end{abstract}


\end{frontmatter}

\section{Introduction}
In 1965, Buchberger developed the method of Gr\"obner bases for solving systems of multivariate non-linear polynomials. Since then, computing power has grown, and improved algorithms developed, but, even now, the method remains impractical for many problems. The main approach to improvement has been to develop algorithms which avoid unnecessary computation culminating in Faug\`ere's F5 algorithm. On the other hand, various methods of representing polynomials have been explored and the impact of data structures on algorithm performance evaluated. A central element of the method is that it is based on the ordering of \emph{power products} (known as \emph{terms} in Faug\`ere's papers)\footnote{In our implementations, we represent monomials as coefficient and power product tuples.}. Orderings that have been used include lexicographic, total degree and variations on these. Prime-based ordering does not appear to have been exploited, and it is the purpose of this paper to explore this case.

The plan of this article is as follows. First, we introduce prime-based ordering as the natural ordering of power products imposed by encoding the indeterminates as distinct prime numbers. Then, we show that this ordering is admissible. We report our implementation of Buchberger's improved algorithm using both total degree ordering and prime-based ordering. Experimental measurements show that significant gains are achieved by using prime-based ordering.

\section{Prime-based ordering: an admissible ordering based on prime numbers}

Total orderings used in Gr\"obner Basis Algorithms are required to be \emph{admissible}. That is they must satisfy the conditions: 
\begin{equation} \forall t : t \not= 1 : 1 < t \end{equation}
\begin{equation} s < t \Rightarrow s \cdot u < t \cdot u \end{equation}

Common admissible total orderings include lexicographical, total degree lexicographical and degree reverse lexicographical. Variations of these and other orderings are possible [\cite{Buch-85}].

An ordering that, to the best of our knowledge, has not been used in implementations of Buchberger's algorithm for Gr\"obner Bases is one based on prime numbers. Given a power product, ${t = x_1^{\alpha_1} x_2^{\alpha_2} \ldots x_n^{\alpha_n}}$,  each indeterminate $x_i$ is mapped to a unique prime number:
\begin{eqnarray} x_1 &\leftrightarrow& 2 \nonumber \\
x_2 &\leftrightarrow& 3 \\
&\ldots& \nonumber \end{eqnarray}
so that ${t \leftrightarrow n_t \in N}$. For example, ${x^3y^2z \leftrightarrow 2^3 3^2 5 = 360}$.

We define
\begin{equation} s < t \Leftrightarrow n_s < n_t \end{equation}
In other words, the ordering of $s$ and $t$ is determined by the natural ordering of the integers $n_s$ and $n_t$. We call the integer $n_t$ the {\em prime image} of $t$.

Now, if all the exponents of $s$ are zero, ${s = 1}$ and ${n_s = 1}$. Similarly, if ${t \not= 1}$, at least one exponent of $t$ is positive, so that ${n_t > 1}$. This establishes condition (1).

Similarly, from (4), ${s < t \Rightarrow n_s < n_t \Rightarrow n_s n_u < n_t n_u \Rightarrow s \cdot u < t \cdot u}$, establishing condition (2).

Hence, the mapping of indeterminates onto a set of prime numbers is an admissible ordering.
Prime-based order is neither a total degree nor a lexicographical ordering. For example, in total degree, ${x^3 > y^2}$; but, in prime-based ordering, ${x^3 \leftrightarrow 8 < 9 \leftrightarrow y^2}$; so ${x^3 < y^2}$. Similarly, ${x^3 < xy }$ in lexicographical ordering; but, ${x^3 \leftrightarrow 8 > 6 \leftrightarrow xy}$, implying ${x^3 > xy }$ in prime-based ordering.

\section{Implementations of Buchberger's Improved Algorithm}

We have developed four versions of Buchberger's improved algorithm [\cite{Buch-85}], which generate a reduced Gr\"obner Basis. They are written in the object oriented programming language Eiffel so that the algorithm is the same but the implementations of power products and coefficients differ in each version. Hence, variations can be readily created and verified not only by comparing output but also by using Eiffel's built-in preconditions and postconditions. Eiffel allows these conditions to be discarded during compilation so that the algorithm can run at full speed. Furthermore, mathematical constructs can be represented as  structures of interconnected objects.

The versions differ in the power-product representation and the integer precision used for the coefficients. In the first version, monomials are represented as structures consisting of a rational coefficient and a power product, ${t = x_1^{\alpha_1} x_2^{\alpha_2} \ldots x_n^{\alpha_n}}$. The coefficient is represented by a pair of 64 bit integers -- numerator and denominator -- and the power product is represented as a string of characters, in expanded form. For example, ${x^3y^2z}$ is stored as the string ``aaabbc''. Power products are compared using total degree in this case. This representation was chosen for simplicity, and, because object oriented programming methods are used, may be changed easily to any other form of representation, such as a vector of powers, ${[3, 2, 1]}$ or as a list. Operations on power products are implemented as iterations or recursions on the underlying data structure. For example, multiplying ${x^3y^2z}$ by ${xyz}$ involves merging the strings ``aaabbc'' and ``abc'' to yield ``aaaabbbcc'', representing ${x^4y^3z^2}$. 

In the second version, primes are used to represent the indeterminates. Power products are ordered by their integer image. This avoids iteration or recursion in the basic operations, which reduce to integer operations. For example, multiplication of power products is integer multiplication, as in  ${x^3y^2z \times xyz \leftrightarrow 360 \times 30 = 10800 = 2^4 3^3 4^2 \leftrightarrow x^4y^3z^2}$. Similarly, the functions of division, lowest common multiple and greatest common divisor also reduce to integer operations. The same is true for Boolean operations such as comparison, equality, divisibility and so on. The prime based implementation of power-products is the same as the original total degree version, but for replacing the routine bodies with the prime based equivalent.

As very large coefficients are often generated, the Gnu multiple precision library, GMP, is used to create the other implementations of the algorithm. This was done by changing the coefficient implementation to use the multiple-precision integer type in the GMP library.

\section{Validation}
Steps were taken to demonstrate that our version of Buchberger's improved algorithm generates Gr\"obner Bases. 
The necessary and sufficient condition to be met by a Gr\"obner Basis, $F$, is :
\begin{equation}
	\forall f_1, f_2 \in F : : NormalForm(F, SPolynomial(f_1, f_2)) = 0 \end{equation}	
where $NormalForm(., .)$ is the normal form or reduction function, and $SPolynomial(., .)$ is leading term cross-reduction function as defined by Definition 6.4 in  [\cite{Buch-85}].	
In the case of a {\em reduced} Gr\"obner Basis, in addition to the above, it is also necessary that
\begin{equation}
	\forall f \in F : : NormalForm(F - \{f\}, f) = f
\end{equation}
  
All sets of polynomials generated by the algorithm were tested for these conditions. It was also verified that each result set reduced the given polynomial set, confirming that the Ideal was unchanged, and that the basis generated by the Gr\"obner Bases package supplied with Maple reduced the result, hence providing independent confirmation.

We found one problem with the implementation of the algorithm as presented in [\cite{Buch-85}]. This was that the set of pairs of polynomials was not updated after a new polynomial was generated by reduction in the $Subalgorithm$ $NewBasis(., ., .)$. To the best of our knowledge, this has not been previously pointed out. When this deficiency was remedied, results proved to be consistent.

\section{Experimental Results and Discussion}

Our timing tests have been performed using the four implementations as shown in Table 1. Examples 1, 2 and 3 are simple manufactured examples, while most of the others are taken from [\cite{Giovini}]. The examples entitled ``Arnold" are taken from [\cite{Arnold}]. 

Care was taken in the ordering of the indeterminates in the case of ``Parametric Curve" as it includes the power product $x^{31}$. This power product can be encoded as $2^{31}$ when using 64-bit integers, but it cannot be encoded as $5^{31}$, which requires 72-bits. 

The main result is that representing the indeterminates by prime numbers and the power products by unique integers significantly reduces the computation time in most cases. In eight cases, the computations complete successfully when the coefficients are encoded as 64-bit integers, and the reduction in time is at least 30\%, or a 40\% speedup. The greatest reduction, (``Gerdt 1"), is 96.8\%, or more than a $30\times$ speedup. As there are more efficient coding schemes than the one used for the total degree implementation, these speedups are on the optimistic side.

A second result is that the number of polynomials may be different. For example, ``Gerdt 1" reduces to 56 polynomials when total-degree ordering is used. When prime-based ordering is used, only 36 polynomials are generated\footnote{Plex ordering, however, generates 26 polynomials for this case.}. However, this is not a general rule, as ``Gerdt 3" generates 21 polynomials for total-degree and 23 for prime-based ordering.  

For some examples, integer overflow problems arise because the coefficients are encoded as pairs of 64-bit integers. To counter this, the Gnu Multiple Precision (GMP) library is used to support coefficients based on very large integers. Using the library allows eleven cases to be compared because they complete successfully with both prime-based and string based power products. A twelfth was completed when garbage collection was turned off.

\begin{center}
	\begin{tabular}{||l||c|c|c||c|c|c||} 
	\multicolumn{7}{c}{\textbf{Table 1. Experimental Results, showing the execution time in ms and}}\\
	\multicolumn{7}{c}{\textbf{the number of polynomials in the generated basis}}
\\\hhline{=======}
	Time (ms)/ & \multicolumn{3}{c||}{64-bit integer coefficients} & \multicolumn{3}{c||}{Multiple-precision integer coefficients} \\ \cline{2-7} 
	\multicolumn{1}{||c||}{Number of }  & \multicolumn{2}{c|}{Power Products} & Reduction & \multicolumn{2}{c|}{Power Products} & Reduction\\ \cline{2-3} \cline{5-6}
	\multicolumn{1}{||r||}{polynomials} & total degree & prime based & { (\%)} & total degree & prime based  &  (\%) \\\hhline{=======}
	Example 1 & 13.22/1 &  9.19/1 & 30.5 & 29.92/1 & 34.09/1 & 12.2 \\\hline
	Example 2 & 2.93/6 & 1.98/6 & 32.4 & 7.32/6 & 6.4/6 & 12.6 \\\hline
	Example 3 &  9.44/6 &  6.35/6 & 32.7 & 24.82/6 & 20.26/6 & 18.4 \\\hline
	Cyclic 4  & 10.73/7 &  5.43/7 & 49.4 & 21.4/7  & 15.54/7 & 27.4 \\\hline
	Cyclic 5  & 12855/20 & \footnotesize{a}      &      & \footnotesize{b}       & 14289/24 & \\\hline
	Gerdt 1   & 345790/56	& 11059/36 & 96.8 & 387004/56 & 14975/36 & 96.1 \\\hline
	Gerdt 2 & 56.8/8 & 4.91/5 & 91.4 & 146.55/8 & 17.74/5 & 87.9 \\\hline
Gerdt 3 & 2693/21 & 1886/23 & 30.0 & \footnotesize {b} & 3217/23 & \\\hline
Gerdt $3^c$ & 4596/21 & 3051/23 & 33.6 & 5584/21 & 4564/23 & 18.3 \\\hline
Arnborg-Lazard & \footnotesize{a} & \footnotesize{a} & & 3042/15 & 2476/11 & 18.6 \\\hline
Parametric Curve   & 420.7/16 & 17.72/10 & 95.8 & 522.5/16 & 35.15/10 & 93.3 \\\hline
Katsura 4 & \footnotesize{a} & \footnotesize{a} &  & 1059/13 & 873/13 & 21.1 \\\hline
Arnold 1 & 153.78/3 & \footnotesize{a} &  & 2276563/3 & 264810/3 & 88.4 \\\hline		
Arnold 2 & \footnotesize{a} & \footnotesize{a} &  & 2919337/2 & 1499222/2 & 48.6 \\\hline		
\end{tabular}
\end{center}
\footnotesize

a. Integer overflow.

b. GMP problem.

c. Garbage collection turned off.

d. Memory exhausted when garbage collection is turned off.

e. Total degree based ordering faster than prime-based ordering.

\normalsize

\hspace{.5in}

With the ``Katsura 4" example, we chanced upon a case in which the algorithm using total-degree ordering was faster than that with prime-based ordering. So, we have carried out a limited investigation of the effect of permuting the relative ordering of the indeterminates of the two polynomial sets, for ``Example 2" with three indeterminates, and for ``Katsura 4" with five indeterminates. We found permutations in which the prime-based ordering was faster than the fastest total-degree ordering in both cases.  

The effects of permuting the relative order of the indeterminates for ``Example 2" are presented in Table 2. In all cases using the 64 bit coefficients, the prime-based ordering version is faster by as much as 46\%. When using GMP, the prime-based ordering is faster by as much as 30\% in all but one case, when it is 16.7\% slower. The fastest computation was the prime-based 64 bit case of 1.98ms when the indeterminate order was $acb$. The ratio of maximum to minimum times is given for each implementation. For example, the ratio of the slowest prime-based 64 bit case to the fastest is 11.

\begin{center}
	\begin{tabular}{||c||c|c|c||c|c|c||} 
	\multicolumn{7}{c}{\textbf{Table 2. Effect of the relative ordering of the indeterminates in ``Example 2" }} \\ \hhline{=======}
	 & \multicolumn{3}{c||}{64-bit integer coefficients} & \multicolumn{3}{c||}{Multiple-precision integer coefficients} \\ \cline{2-7}
	\multicolumn{1}{||r||} {}  & \multicolumn{2}{c|}{Power Products} & Reduction & \multicolumn{2}{c|}{Power Products} & Reduction\\ \cline{2-3} \cline{5-6}
	\multicolumn{1}{||c||}{Order} & total degree & prime based & { (\%)} & total degree & prime based  &  (\%) \\\hhline{=======}
	$abc$ & 15.85 & 10.56 & 33.6 & 38.22 & 33.57 & 12.2 \\\hline
	$acb$ & 2.93 & 1.98 & 32.4 & 7.32 & 6.4 & 12.6 \\\hline
	$bac$ & 24.06 & 18.76 & 22.0 & 60.95 & 58.73 & 3.6 \\\hline
	$bca$ & 41.48 & 22.29 & 46.3 & 125.42 & 88.22 & 29.7 \\\hline
	$cab$ & 3.71 & 2.62 & 29.4 & 9.42 & 8.36 & 11.3 \\\hline
	$cba$ & 15.08 & 13.15 & 12.8 & 37.50 & 45.01 & $\textbf{(16.7)}^a$ \\\hhline{=======}
	max/min & 14 & 11 && 17 & 14 &\\\hhline{=======}

\end{tabular}
\end{center}

\footnotesize 
 
a. Total degree based ordering faster than prime-based ordering.

\normalsize

\hspace{.5in}

In the case of ``Katsura 4", there are five indeterminates, and 120 orders to consider. Only the multiple precision implementations work, as the 64-bit coefficient implementations overflow. The fastest case is prime based, with a duration 17.6\% less than fastest total degree based case. The ratio of the slowest to fastest is 1.9 for the total degree based implementation and 3 for the prime-based implementation.

In the ``Gerdt 3" example and some other examples, using the GMP library triggered a fault which caused the program to crash. This crash occurs when objects are collected by the garbage collector while running or when the program terminates. This is believed to be a problem in the interface between Eiffel and the GMP library rather than GPM itself. Turning off garbage collection avoided the fault, but proved costly, as it is faster to re-allocate memory internally than by repeatedly requesting additional memory from the operating system. Secondly, with garbage collection off, if memory is exhausted, the computation becomes disk-bound; these cases were abandoned.

\section{Conclusions}

Prime-based ordering, based on ordering power products by encoding the indeterminates as prime numbers and using the natural number order, is an admissible ordering. Prime-based ordering is not a lexicographical or total degree ordering. Implementations of this ordering reduce power product operations to integer operations. 

Several versions of Buchberger's improved algorithm have been developed and tested. Each result has been verified to satisfy the necessary and sufficient conditions to be a reduced Gr\"obner Basis. Resulting bases also reduce their respective given polynomial sets, confirming that the Ideal is correctly preserved. They were also shown to be reduced by bases generated using the Gr\"obner package in Maple.

Duration reductions measured using the improved Buchberger algorithm range from 30\% (40\% speedup) to 96.8\% ($30\times$ speedup). 

The number of polynomials can differ according to the ordering scheme. For example, in the ``Gerdt 1" case, prime-based ordering generates a Gr\"obner basis with 36 polynomials, whereas the total degree result has 56 polynomials.

Finally, we have also explored the effect of permuting the indeterminates in some examples, and have found that the duration of computation varies significantly. In these examples, the fastest case was always using the prime-based ordering. We are continuing to investigate this matter together with issues associated with large coefficient size.

\bibliographystyle{elsart-harv}


\appendix
\section{Polynomial Sets}

\subsection{Example 1}
$\begin{array}{l}
c^4 - 6ac^3 + 13a^2c^2 - 12a^3c + 4a^4 \\
b^2 - 2ab - 2bc + a^2 + 2ac + c^2 \\
a^2 + 4a + 3 \\
ac^3 - 3c^2a^2 + 3ca^3 - a^4 + c^3 - 3c^2a + 3ca^2 - a^3 + ac - 2a^2 +3c - 6a \\
5abc^4+ 3ab^2 + a + 1
\end{array}$

\subsection{Example 2}

$\begin{array}{l}
	c^4 - 6ac^3 + 13a^2c^2 - 12a^3c + 4a^4 \\
	b^2 - 2ab - 2bc + a^2 + 2ac + c^2 \\
	a^2 + 4a + 3 \\
	ac^3 - 3c^2a^2 + 3ca^3 - a^4 + c^3 - 3c^2a + 3ca^2 - a^3 + ac - 2a^2 +3c - 6a
\end{array}$

\subsection{Example 3}

$\begin{array}{l}
	c^4 - 6ac^3 + 13a^2c^2 - 12a^3c + 4a^4 \\
	a^2 + 4a + 3 \\
	ac^3 - 3c^2a^2 + 3ca^3 - a^4 + c^3 - 3c^2a + 3ca^2 - a^3 + ac - 2a^2 +3c - 6a \\
	5abc^4+ 3ab^2 + a + 1 
\end{array}$

\subsection{Cyclic 4}
$\begin{array}{l}
x + y + z + t \\
xy + yz + zt + tx \\
xyz + yzt + ztx + txy \\
xyzt - 1
\end{array}$

\subsection{Cyclic 5}
$\begin{array}{l}
x + y + z + t + u \\
xy + yz + zt + tu + ux \\
xyz + yzt + ztu + tux + uxy \\
xyzt + yztu + ztux + tuxy + uxyz \\
xyztu - 1
\end{array}$

\subsection{Gerdt 1}
$\begin{array}{l}
yw-1/2zw + tw \\
-2/7uw^2+10/7vw^2-20/7w^3+tu-5tv+0tw \\
2/7yw^2-2/7zw^2+6/7tw^2-yt+zt-3t^2 \\
-2v^3+4uvw+5v^2w-6uw^2-7vw^2+15w^3+42yv \\
-14zv-63yw+21zw-42tw+147x \\
-9/7uw^3+45/7vw^3-135/7w^4+2zv^2-2tv^2-4zuw+10tuw-2zvw-28tvw \\  \hspace{.5in} +4zw^2+86tw^2-42yz +14z^2+42yt-14zt-21xu+105xv-315xw \\
6/7yw^3-9/7zw^3+36/7tw^3-2xv^2-4ytw+6ztw-24t^2w+4xuw+2xvw \\  \hspace{.5in} -4xw^2+56xy-35xz+84xt \\
2uvw-6v^2w-uw^2+13vw^2-5w^3+14yw-28tw \\
u^2w-3uvw+5uw^2+14yw-28tw \\
-2zuw-2tuw+4yvw+6zvw-2tvw-16yw^2-10zw^2+22tw^2+42xw \\
28/3yuw+8/3zuw-20/3tuw-88/3yvw-8zvw+68/3tvw+52yw^2+40/3zw^2 \\  \hspace{.5in} -44tw^2-84xw \\
-4yzw+10ytw+8ztw-20t^2w+12xuw-30xvw+15xw^2 \\
-y^2w+1/2yzw+ytw-ztw+2t^2w-3xuw+6xvw-3xw^2 \\
8xyw-4xzw+8xtw
\end{array}$

\subsection{Gerdt 2}
$\begin{array}{l}
35y^4 - 30xy^2 - 210y^2z + 3x^2 + 30xz - 105z^2 + 140yt - 21u \\
5xy^3 - 140y^3z - 3x^2y + 45xyz - 420yz^2 + 210y^2t -25xt + 70zt + 126yu
\end{array}$

\subsection{Gerdt 3}
$\begin{array}{l}
6xy^2t - x^2zt - 6xyzt + 3xz^2t - 2z^3t - 6xy^2 + 6xyz - 2xz^2 \\
-63xy^2t^2 + 9x^2zt^2 + 63xyzt^2 + 18y^2zt^2 - 27xz^2t^2 -18yz^2t^2 + 18z^3t^2 + 78xy^2t \\  \hspace{.5in}  - 78xyzt - 18y^2zt +24xz^2t + 18yz^2t - 9z^3t - 15xy^2 + 15xyz - 5xz^2 \\
18x^2y^2t - 3x^3zt - 18x^2yzt + 12xy^2zt + 5x^2z^2t - 12xyz^2t +6xz^3t - 8z^4t - 18x^2y^2 \\  \hspace{.5in}  + 18x^2yz -12xy^2z - 4x^2z^2 + 12xyz^2 - 6xz^3 \\
-x^2yt + 3xy^2t + 10y^3t - 15y^2zt + 3yz^2t - 3xy^2 - 10y^3 +xyz + 15y^2z - 5yz^2
\end{array}$

\subsection{Arnborg-Lazard}
$\begin{array}{l}
x^2yz + xy^2z + xyz^2 + xyz + xy + xz + yz \\
x^2y^2z + x^2yz + xy^2z^2 + xyz + x + yz + z \\
x^2y^2z^2 + x^2y^2z + xy^2z + xyz + xz + z + 1
\end{array}$

\subsection{Parametric Curve}
$\begin{array}{l}
x^31 - x^6 - x - y \\
x^8 - z \\
x^10 - t
\end{array}$

\subsection{Katsura 4}
$\begin{array}{l}
2x^2 + 2y^2 + 2z^2 + 2t^2 + u^2 - u \\
xy + 2yz + 2zt + 2tu - t \\
2xz + 2yt + t^2 + 2zu - z \\
2xt + 2zt + 2yu - y \\
2x + 2y +2z +2t +u - 1
\end{array}$

\subsection{Arnold 1}
$\begin{array}{l}
8x^2y^2+5xy^3+3x^3z+x^2yz \\
x^5+2y^3z^2+13y^2z^3+5yz^4 \\
8x^3+12y^3+xz^2+3 \\
7x^2y^4+18xy^3z^2+y^3z^3
\end{array}$

\subsection{Arnold 2}
$\begin{array}{l}
2xy^4z^2 + x^3y^2z - x^2y^3z + 2xyz^2 + 7y^3 + 7 \\
2x^2y^4z + x^2yz^2 - xy^2z^2 + 2x^2yz - 12x + 12y \\
2y^5z + x^2y^2z - xy^3z - xy^3 + y^4 + 2y^2z \\
3xy^4z^3 + x^2y^2z - xy^3z + 4y^3z^2 + 3xyz^3 + 4z^2 - x + y
\end{array}$

\end{document}